*Research Article*

# Methods to Estimate Surface Roughness Length for Offshore Wind Energy

## Maryam Golbazi and Cristina L. Archer

*College of Earth, Ocean, and Environment, University of Delaware, Newark, DE 19716, USA*

Correspondence should be addressed to Cristina L. Archer; carcher@udel.edu

The northeastern coast of the U.S. is projected to expand its offshore wind capacity from the existing 30 MW to over 22 GW in the next decade, yet, only a few wind measurements are available in the region and none at hub height (around 100 m today); thus, extrapolations are needed to estimate wind speed as a function of height. A common method is the log-law, which is based on surface roughness length ($z_0$). No reliable estimates of $z_0$ for the region have been presented in the literature. Here, we fill this knowledge gap using two field campaigns that were conducted in the Nantucket Sound at the Cape Wind (CW) platform: the 2003–2009 "CW Historical", which collected wind measurements on a meteorological tower at three levels (20, 41, and 60 m AMSL) with sonic and cup/vane anemometers, and the 2013–2014 IMPOWR (Improving the Mapping and Prediction of Offshore Wind Resources), which collected high-frequency wind and flux measurements at 12 m AMSL. We tested three different methods to calculate $z_0$: (1) analytical method, dependent on friction velocity $u_*$ and a stability function $\psi$; (2) the Charnock relationship between $z_0$ and $u_*$; and (3) a statistical method based on wind speed observed at the three levels. The first two methods are physical, whereas the statistical method is purely mathematical. Comparing mean and median of $z_0$, we find that the median is a more robust statistics because the mean varies by over four orders of magnitude across the three methods and the two campaigns. In general, the median $z_0$ exhibits little seasonal variability and a weak dependency on atmospheric stability, which was predominantly unstable (54–67%). With the goal of providing the most accurate estimates of wind speed near the hub height of modern turbines, the statistical method, despite delivering unrealistic $z_0$ values at times, gives the best estimates of 60 m winds, even when the 5 m wind speed from a nearby buoy is used as the reference. The unrealistic $z_0$ values are caused by nonmonotonic wind speed profiles, occurring about 41% of the time, and should not be rejected because they produce realistic fits. Furthermore, the statistical method outperforms the other two even though it does not need any stability information. In summary, if wind speed data from multiple levels are available, as is the case with vertically pointing floating lidar and meteorological towers, the statistical method is recommended, regardless of the seemingly unrealistic $z_0$ values at times. If multilevel wind speeds are not available but advanced sonic anemometry is available at one level, the analytical method is recommended over Charnock's. Lastly, if a single, constant value of $z_0$ is sought after to characterize the region, we recommend the median from the statistical method, i.e., $6.09 \times 10^{-3}$ m, which is typical of rough seas.

## 1. Introduction

Offshore wind farms have the potential to become a major and national source of electricity in the U.S., as winds are generally stronger and steadier at offshore than inland areas [1]. Reports of the Department of Energy (DOE) in the U.S. suggest a potential power capacity of 2,000 gigawatts (GW) per year from offshore wind sites along the coasts of the U.S. and big lakes. This amount is about two times the combined energy generation from all electric power plants in the whole country [2]. The northeastern coastal waters of the U.S. are particularly favorable for future offshore wind farm development in the U.S. [3], partly because these areas already pay the highest electric utility rates in the nation [1]. According to the DOE, the East Coast may be capable of providing its own electricity from offshore wind better than from any other form of energy generation [2]. The U.S. first-ever offshore wind farm was constructed off the coast of Rhode Island with a single 30-megawatt (MW) project in 2016. It includes five 6-MW turbines off of Block Island,

Rhode Island. The reports on the Block Island project show an increase in tourism, a decrease in energy price, and many more advantages (https://www.awea.org/policy-and-issues/u-s-offshore-wind). Planned offshore wind installations in the U.S. will total 22 GW by 2030 and 86 GW by 2050, mostly off the East Coast [2]. This study analyses wind data available from the Nantucket Sound area, located in the northeastern coast of the U.S. offshore of Massachusetts.

Since the power production of a wind turbine is related to the cube of the hub-height wind speed [4], accurate measurements on wind velocities at or near hub height will lead to precise predictions of energy production. Aerodynamic surface roughness length $z_0$ is a critical parameter to extrapolate the observed wind speeds from one fixed height, typically near the surface, to another specified height, for example at hub height, and therefore to estimate the vertical wind profiles [5]. The main goal of this study is to explore the best estimate of surface roughness length for the northeastern coast of the United States based on the data available and to investigate the best fitting profiles of offshore wind velocities near hub height for offshore wind development.

Surface roughness length in the offshore marine environment is generally lower than that inland and mostly depends on the wave field properties: the higher the waves, the higher the ocean surface roughness length [6]. However, there are exceptions. Frank et al. [7] report that in lower wind speeds, sea surface roughness increases rather than decreasing. They also show that in near-neutral and stable atmospheric conditions, the wind shear at higher elevations above the water is underestimated. Similarly, Archer et al. [8] find that atmospheric stability is a critical factor in designing wind farms due to its impact on wind shear in the atmospheric boundary layer (ABL), which has an influence on hub-height wind speed and therefore on power production. Kim et al. [9] conclude that the effects of an inaccurate surface roughness length are not significant on wind speed prediction; by contrast, using wind measurements at a higher level results in a decrease in the prediction errors.

Some researchers believe that a constant value for surface roughness length is sufficient, since it results in deviations of the wind speed profile from observations at higher elevations, where the profile is relatively flat [10]. For instance, the Wind Speed Estimation program (WAsP) assumes a value of $z_0 = 0.0002$ m over the ocean [11], and Garratt [12] mentions that surface roughness length is a small fraction of the height of the bumps on the ocean surface, about one tenth of them. A study by Lange et al. [10] shows that various methods for obtaining surface roughness length do not make a significant difference on its value.

On the other hand, surface roughness length is considered an important value in wind speed calculations by other researchers. For instance, Ueno and Deushi [13] take an approach that considers the wave characteristics to calculate roughness length. Similarly, Wu [14] concludes that surface roughness length in low-wind ranges decreases with increasing wind speed and in high-wind ranges increases with wind speed. Wever [15] conducted simulations with a conceptual ABL model to discover that 70% of the wind speed trends can be contributed to by the aerodynamic surface roughness length. Jimenez and Dudhia [16] discuss the importance of water depth on the surface roughness calculation and believe that there is a major wind speed bias when comparing the same model results on an open ocean versus a shallow water site, due to the higher surface drag over shallow water than that on the open ocean. A study by Donelan [17] investigates the impacts of surface roughness length and wind speed on each other in the marine environment and show that the ocean surface does not get any rougher when the wind speed on water exceeds 33 m/s.

When calculating $z_0$, an uncertainty arises about the so-called displacement height [18], which is a correction to the height above ground to include the effect of canopies and other obstacles that are located upwind of a site and that will "raise" the point at which the extrapolated wind profile would get to zero, even though the surface roughness length remains unaffected. Kim et al. [9] find that the error decreases when considering a displacement height and conclude that, in areas where sea surface level varies significantly from the average sea level, calculations should be done by considering these changes. Khan et al. [19] observe a slight dependence of the ratio of wind speed from two different heights ($U_2/U_1$) on tidal variation. However, Elkintone et al. [20] mention that the tidal variations are not found to have significant impacts on wind speed profiles. They also attempt to estimate the surface roughness length using data from two heights in the Nantucket Sound area and recommend a $z_0$ value of 0.1 m. This value seems higher than what is expected for surface roughness length in an offshore area. The current study will clarify the reasons of such high value and provide alternative recommendations.

Three methods existing in the literature—analytical, statistical, and Charnock—are used here to calculate the surface roughness length in the Nantucket Sound area offshore of the East Coast of the U.S. (Figure 1). The first method is an analytic solution to the problem with a focus on stability information. The second method uses the Charnock relation. The third is a statistical method with an emphasis on error minimization and the best fit of the vertical wind profile. The three data sets used in this study are described in the next Section 2, the details of the three methods in Section 3, and the results in Section 4.

## 2. Data

Three long-term observational data sets are used in this study. The first data set comes from the 2003–2009 Cape Wind (CW) campaign at Nantucket Sound (Figure 1), which hosted a meteorological tower with three levels of measurements (20, 41, and 60 m AMSL) with both 3D sonic and cup/vane anemometers. However, due to malfunctioning of the instruments after 2007, only the data for the years 2003–2007 were used in this research. The Cape Wind data set, hereafter referred to as "CW Historical," consists of 10-minute observations of meteorological variables such as wind speed and direction at the three levels, heat and momentum fluxes, temperature, etc. The original data set included 227,353 data points. However, a number of data points were missing or were cleaned up to remove unreasonable values, as described in [21]. Therefore, 214,458 valid data points were retained for this study (Table 1).

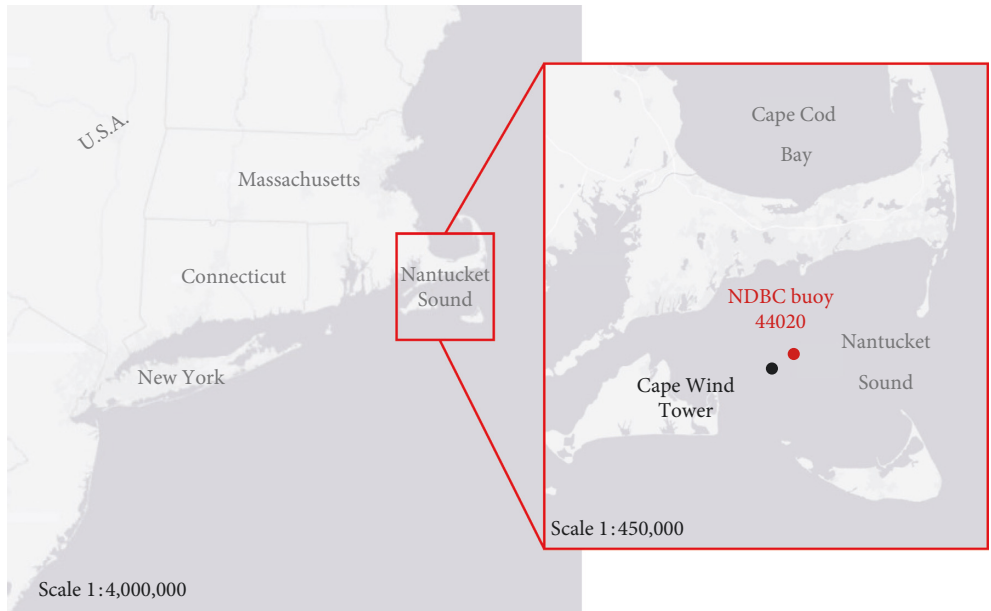

Figure 1: Location of the Cape Wind tower and the National Data Buoy Center (NDBC) buoy 44020 in the Nantucket Sound area off the coast of Massachusetts.

Data from the Improving the Mapping and Prediction of Offshore Wind Resources (IMPOWR) campaign were collected from two sonic anemometers located on the CW platform at 12 m AMSL, as described in [4]. The two anemometers were placed across from each other to collect the data from opposite directions. The number of data points for IMPOWR was 105,112, but only about 19,633 were retained after data cleaning and included enough information to calculate the $z_0$ values (Table 1).

A third data set was obtained from the National Data Buoy Center [22]. The data set included observations from the 44020 buoy located in the middle of the Nantucket Sound (Figure 1), with a mast at 5 m AMSL. It consists of 10-minute wind speed data for the years 2003 to 2007, consistent with the CW Historical data set. The number of data points in the NDBC data set was about 350,000. However, only 202,830 were retained after the data cleaning procedure (Table 1).

## 3. Methods

Atmospheric stability refers to the tendency of the atmosphere to enhance or suppress vertical motion, in which cases an atmospheric layer is called unstable or stable, respectively. When vertical motion is neither suppressed nor enhanced, the layer is called neutral [23]. Heat fluxes are upward in unstable, downward in stable, and zero in neutral conditions and generally turbulence is increasingly higher from stable to unstable conditions [8]. Atmospheric stability also impacts the shape of the mean wind speed profile, wind direction, and turbulence around a wind turbine [24]. Thus, it influences the surface roughness length.

A common parameter to estimate atmospheric stability is the Obukhov length $L$, which represents the lowest height above the ground at which turbulence production by buoyancy dominates over that by mechanical effects, such as shear and friction [25]. The actual calculation of $L$, however, depends on the type of measurements available. Given that 3D sonic anemometers were available in both IMPOWR and CW Historical, the following equation for $L$ was used [8]:

$$L = -\frac{u_*^3}{\kappa\left(g/T_S\right)\overline{w'T_S'}},$$ (1)

where $\overline{w'T_S'}$ is the sonic heat flux (positive for upward and negative for downward fluxes), $\kappa$ is the von Kármán constant ($\kappa = 0.41$), $g$ is gravity, and $T_S$ is the near-surface sonic temperature. The value of $L$ is infinite in a neutral atmosphere inasmuch as thermal exchanges are absent by definition. Conversely, in both stable and unstable conditions, buoyancy is the driving force and therefore, the magnitude of $L$ is small, indicating that shear and friction are important only close to the surface, and the sign is positive and negative, respectively.

Table 1: Details about the two field campaigns conducted at the Cape Wind tower in Nantucket Sound and about the NDBC data set used in this study.

| Campaign | Years | Reference level | Other levels | Level of $u_*$ | N. of valid data |
|---|---|---|---|---|---|
| CW Historical | 2003–2007 | 20 m | 41 m, 60 m | 20 m | 214,458 |
| IMPOWR | 2013-2014 | 12 m | — | 12 m | 19,633 |
| NDBC | 2003–2007 | 5 m | — | — | 202,830 |

To incorporate the effect of the measurement height, the stability parameter $\zeta$, a function of $L$, is used here:

$$\zeta = \frac{z_S}{L},$$ (2)

where $z_S$ is the height of the sonic anemometer (12 m and 20 m for IMPOWR and CW Historical, respectively).

Based on the stability parameter, the atmosphere stability is evaluated as follows:

(i) Stable: $\zeta > (z_S/500)$

(ii) Unstable: $\zeta < (-(z_S/500))$

(iii) Neutral: $|\zeta| \leq (z_S/500)$

In the neutral atmospheric boundary layer, the wind velocity profile is expected to be logarithmic. Two properties are relevant in this case: friction velocity $u_*$ and $z_0$. Friction velocity is related to shear stress at the surface [26] and can be calculated as

$$u_* = \sqrt[4]{\overline{u'w'}^2 + \overline{v'w'}^2}, \tag{3}$$

where $\overline{u'w'}$ and $\overline{v'w'}$ are the covariances of the surface momentum fluxes.

The logarithmic wind speed profile, also called the "log-law," takes the following form:

$$U(z) = \frac{u_*}{\kappa}\left[\log\left(\frac{z-d}{z_0}\right) - \psi\right], \tag{4}$$

where $U(z)$ is the wind speed at height $z$ AMSL, $d$ is the displacement height, and $\psi$ is the stability correction. Since our case study is a marine environment with no canopies or tall obstacles, the displacement height is not applicable and is set to zero ($d = 0$). The stability correction $\psi$ depends on $\zeta$ and is different for neutral, stable, and unstable atmospheric conditions as follows:

$$\psi = \begin{cases} 0, & \text{if neutral,} \\ 2\log\left(\frac{1+x}{2}\right) + \log\left(\frac{1+x^2}{2}\right) - 2\arctan(x) + \frac{\pi}{2}, & \text{if unstable,} \\ -5\zeta, & \text{if stable,} \end{cases} \tag{5}$$

where $x = (1 - \zeta)^{1/4}$.

The log-law in Equation (4) requires information on stability to calculate $\psi$, as well as $u_*$. Three-dimensional (3D) sonic anemometers can provide both; however, they are sophisticated and expensive instruments and thus relatively rare. If 3D sonic anemometers are not available, then a common approach is to set the stability parameter to zero, thus assuming that the atmosphere is neutral, which is considered as a valid assumption in the marine boundary layer. To eliminate the dependency on $u_*$, wind speed measurements from cup or 2D sonic anemometers at a reference height $z_R$ (i.e., the height at which the cup or 2D sonic anemometer is mounted) can be used in this alternative form of the log-law:

$$U(z) = U(z_R)\frac{\log(z/z_0)}{\log(z_R/z_0)}, \tag{6}$$

which is obtained by imposing Equation (4) through $z_R$ with wind speed equal to $U(z_R)$ and through a generic level $z$ with wind speed equal to $U(z)$, then taking the ratio of the two to effectively eliminate $u_*$. We will refer to this second form of the log-law in Equation (6) as "simplified log-law" since it does not require any flux measurements or stability information. To evaluate $z_0$, three methods are proposed next, based on the data available from the two field campaigns.

As discussed in Archer et al. [8], the data from IMPOWR include a disproportionately high number of measurements in the month of April (over 30%), when the average wind speed is high (~8 m/s at 20 m). On the other hand, there are no data available for the month of July and the number of observations available for summer, when the average wind speed is lowest (~6.5 m/s at 20 m), were much fewer in comparison to the other seasons (Table 2). As a result, the simple mean wind speed at IMPOWR would be biased high because the high-wind speeds in April would be noticeably overweighted and the summer low-wind speeds underweighted. To address this issue, the average wind speed values for both campaigns were calculated giving an equal weight to all months. Monthly averages $\overline{U}_n$ were calculated first for each month $n$ and then all the months were weighted equally in calculating the total mean value as follows:

$$\overline{U} = \frac{\sum_{n=1}^{n=12}\overline{U}_n}{12}. \tag{7}$$

For IMPOWR, the missing mean wind speed for the month of July was estimated as the average of the June and August means, based on Figure 2(a) in [8].

### 3.1. Analytical Method.

With the analytical method, Equation (4) can be rewritten for $z = z_R$ and solved for $z_0$ as follows:

$$z_0 = \frac{z_R}{\exp\left((\kappa U(z_R))/u_* + \psi\right)}. \tag{8}$$

The analytical method is a physical method and is based on physical properties measured at a given location, namely, wind speed at the reference height, friction velocity, and atmospheric stability (Table 3). By contrast, the other two methods that will be discussed in this section do not depend on as many physical variables as the analytical method.

The analytical method was applied to 100,394 20 m measurements in the CW Historical data set and to 19,633 12 m measurements in the IMPOWR data set. The 20 m reference height was selected from the CW Historical data set because it was the closest height to the sea surface level. Following the fact that the analytical method is a physics-based method, the resulting values of $z_0$ are supposed to have a physical meaning as well and therefore nonphysical, unrealistic values should not be retained. Values of $z_0$ greater than 1 m, the average value for urban areas and big cities, were removed. As a result, 18,764 and 99,309 valid $z_0$ values for IMPOWR and CW, respectively, were retained.

Table 2: Mean and standard deviation of wind speed (by season and overall) measured from buoy 44020 during 2003–2007 at 5 m, the IMPOWR campaign during 2013–2014 at 12 m, and the CW tower during 2003–2007 at three different heights. All heights are in meters above mean sea level.

| | Buoy | IMPOWR | CW Historical | | |
| --- | --- | --- | --- | --- | --- |
| | 5 m | 12 m | 20 m | 41 m | 60 m |
| ALL | | | | | |
| Mean (m/s) | 6.52 | 7.39 | 7.84 | 8.43 | 8.82 |
| Standard deviation (m/s) | 3.58 | 3.25 | 3.66 | 3.92 | 4.12 |
| Count | 202,830 | 19,633 | 214,458 | 214,458 | 214,458 |
| DJF | | | | | |
| Mean (m/s) | 8.73 | 7.95 | 9.01 | 8.20 | 8.58 |
| Standard deviation (m/s) | 3.76 | 3.35 | 4.09 | 4.30 | 4.45 |
| Count | 48,972 | 3,353 | 50,592 | 50,592 | 50,592 |
| MAM | | | | | |
| Mean (m/s) | 6.46 | 7.67 | 7.79 | 8.51 | 9.06 |
| Standard deviation (m/s) | 3.35 | 3.33 | 3.60 | 3.91 | 4.17 |
| Count | 47,701 | 9,677 | 50,974 | 50,974 | 50,974 |
| JJA | | | | | |
| Mean (m/s) | 4.40 | 6.51 | 6.46 | 6.95 | 7.48 |
| Standard deviation (m/s) | 2.16 | 2.25 | 2.50 | 2.80 | 3.10 |
| Count | 52,331 | 2,399 | 54,567 | 54,567 | 54,567 |
| SON | | | | | |
| Mean (m/s) | 6.48 | 7.97 | 7.81 | 8.28 | 8.65 |
| Standard deviation (m/s) | 3.48 | 3.21 | 3.80 | 4.08 | 4.22 |
| Count | 53,826 | 4,203 | 58,325 | 58,325 | 58,325 |

### 3.2. Charnock.

A common method in the literature to parameterize $z_0$ in the marine environment is the Charnock relation [27]:

$$z_0 = \alpha \frac{u_*^2}{g}, \qquad (9)$$

where $\alpha$ is the Charnock parameter, which mostly depends on the wave age [10, 13]. Many experimental studies exist for calculating $\alpha$, including fitting field data [28]. For instance, Frank et al. [7] suggested a value of 0.018 for coastal areas and 0.011 for open seas. Lange et al. [10] used a value of 0.0185 for their calculations. Amongst all the values used for $\alpha$, however, a review by Garratt [29] shows that the average value in the literature is 0.0144 and therefore, this will be the value selected for $\alpha$ in this study, as done also by Van Wijk et al. [30].

To calculate surface roughness length via the Charnock equation, only friction velocity is needed, and no stability information is used directly (Equation (9)). However, since a 3D sonic anemometer can provide both $u_*$ and stability information, either Equation (4) or (6) can be used to calculate $U(z)$ once the estimate for $z_0$ from Charnock is obtained (Table 3).

The number of valid $z_0$ values obtained from the Charnock equation was 18,819 and 100,393 for IMPOWR and CW Historical, respectively. The $z_0$ values calculated from the Charnock method were all in the range of $10^{-6}$–$10^{-2}$ m; thus, no thresholds were enforced.

### 3.3. Statistical Method.

A statistical methodology to extrapolate wind speed at a given elevation was provided by Archer and Jacobson [31, 32], based on the least-square-error approach. The equation for $z_0$ that would give the minimum (squared) error is

$$\ln(z_0) = \frac{U(z_R)\left\{\sum\left[\ln(z_i)\right]^2 - \ln(z_R)\sum\ln(z_i)\right\} - \ln(z_R)\sum\left[U_i\ln(z_i/z_R)\right]}{\left\{U(z_R)\sum\ln(z_i) - \sum\left[U_i\ln(z_i/z_R)\right] - NU(z_R)\ln(z_R)\right\}}, \qquad (10)$$

where $i$ is the index for the vertical levels (at CW $i = 1$, 2, or 3), and $U_i$ is the wind speed at height $z_i$.

As mentioned before, this method is purely mathematical and requires no information about stability (Table 3). Equation (10) only requires the wind speeds observed at various heights. As such, the estimates for $z_0$ are not expected to be physical or realistic when the statistical method is used because they will compensate for the lack of stability information with possibly unrealistic values of $z_0$ that yet give accurate estimates of wind speed. No thresholds were therefore applied to the estimates of $z_0$ from the statistical method. The number of valid $z_0$ values obtained from the statistical method was 207,676, which is significantly higher in comparison to the other two methods discussed previously, due to the fact that no stability information was required. The only values that were rejected for the statistical method were the singularity points that resulted in a zero denominator in Equation (10), about 7000.

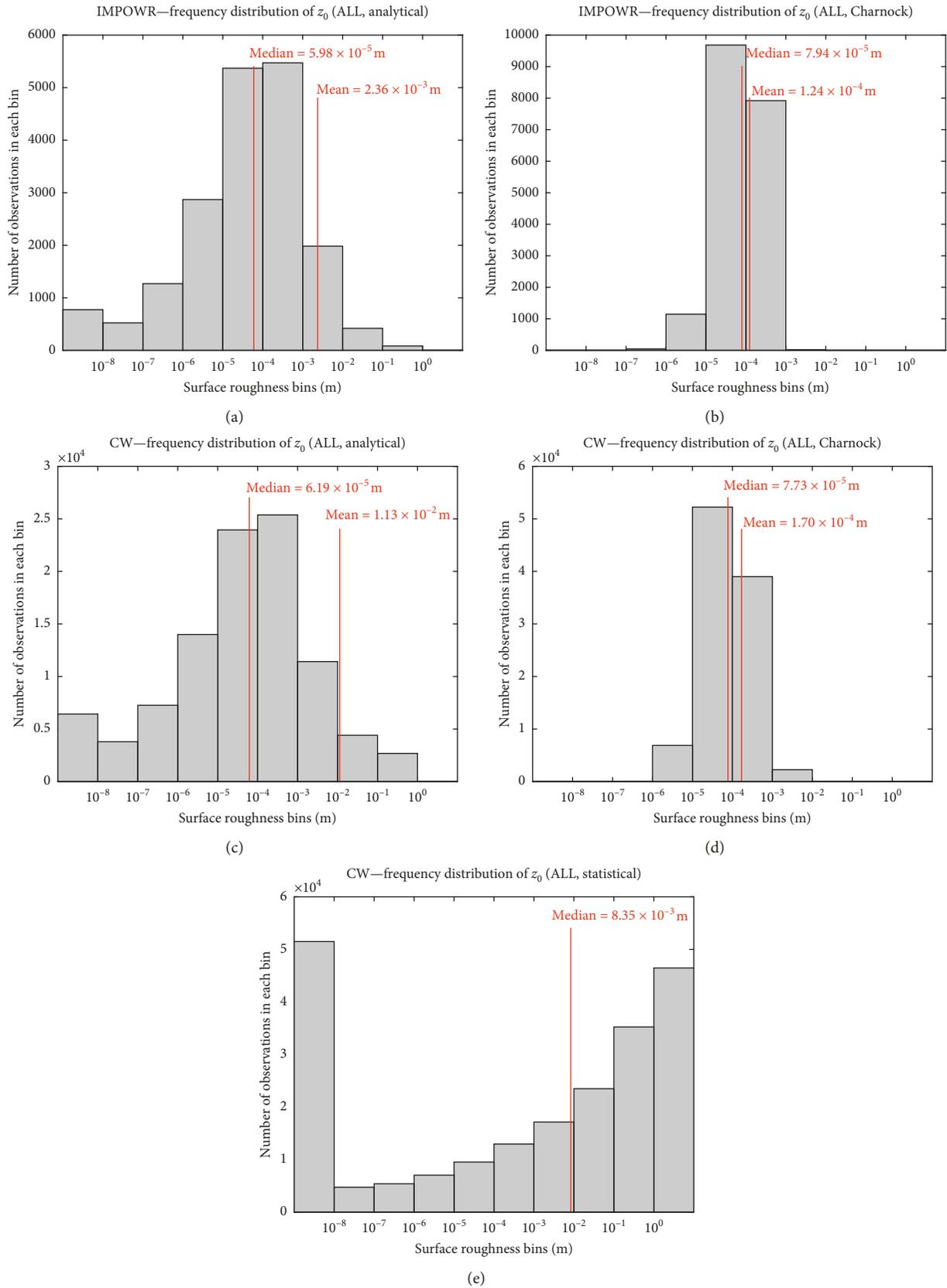

FIGURE 2: Frequency distribution of surface roughness length from the 2013–2014 IMPOWR campaign with two sonic anemometers located at 12 m AMSL and (a) the analytical method and (b) the Charnock method and the 2003–2007 CW Historical campaign with (c) the analytical method using a sonic anemometer located at 20 m AMSL, (d) the Charnock method using a sonic anemometer located at 20 m AMSL, and (e) the statistical method using pairs of sonic and cup anemometers at 20, 41, and 60 m AMSL. The mean and median $z_0$ are displayed on each figure. The $x$-axis is logarithmic and the $y$-axis has different maxima in each figure. Note that the statistical method is not intended to give realistic values of surface roughness length; thus, its distribution is shown here in (e) just for completeness.

Table 3: Information used to calculate $z_0$ with each of the three methods and associated equation(s) to calculate the vertical wind speed profile $U(z)$.

| Method to estimate $z_0$ | Information available to calculate $z_0$ | | | | Equation to calculate $U(z)$ | |
| | One 3D sonic anemometer | | | Sonic/cup at $\geq 3$ levels | Log-law | Simplified log-law |
| | Stability | $u_*$ | Wind speed | Wind speed | (Equation (4)) | (Equation (6)) |
| Analytical | × | × | × | | × | × |
| Charnock | | × | | | × | × |
| Statistical | | | | × | | × |

## 4. Results

### 4.1. Observed Surface Roughness Properties and Statistics.
The frequency distribution of $z_0$ obtained from the three methods is provided in Figure 2 for both campaigns, and the main statistics are displayed in Table 4. In general, the analytical and Charnock methods give results that are consistent with expectations for the marine environment, whereas the statistical method gives unrealistic $z_0$ values at times, and therefore, it will be discussed separately in Section 4.2.

The median $z_0$ values are generally more consistent than the mean. For example, the mean $z_0$ values from the analytical and Charnock methods are $1.13 \times 10^{-2}$ m and $1.70 \times 10^{-4}$ m for the CW Historical and $2.36 \times 10^{-3}$ m and $1.23 \times 10^{-4}$ m for IMPOWR, respectively (Table 4). The median $z_0$ values from the analytical and Charnock methods are $6.19 \times 10^{-5}$ m and $7.73 \times 10^{-5}$ m for the CW Historical and $5.98 \times 10^{-5}$ m and $7.94 \times 10^{-5}$ m for IMPOWR, respectively. Thus, the analytical and Charnock methods give consistent results between the two campaigns, although the mean $z_0$ from the analytical method at CW Historical is larger than that from IMPOWR. The $z_0$ distribution from the Charnock method is narrower around the mean and the median (i.e., lower standard deviation) relative to that from the analytical method (Figure 2), while the medians are consistent for both methods and both campaigns (Table 4).

Next, we explore the impacts of physical phenomena such as atmospheric stability and seasonality on the values of offshore surface roughness length. Figure 3 shows the frequency distribution of $z_0$ from the analytical method for both campaigns for neutral, stable, and unstable cases separately. The distributions are similar in shape and median values, although a higher frequency of low $z_0$ values is found for neutral conditions during the CW Historical campaign (first bin in Figure 3(d)). Again, the medians are more consistent among the two campaigns than the means. Overall, the $z_0$ values in this offshore area do not vary significantly as a result of atmospheric stability. Therefore, it can be concluded that $z_0$ is relatively insensitive to atmospheric stability.

Note that $z_0$ is expected to be independent of stability at inland locations. From Equation (4), atmospheric stability impacts the vertical wind speed profile through the stability correction $\psi$, not through $z_0$. At offshore locations, by contrast, $z_0$ is affected by the waves, which are related to the wind near the surface, which in turn is affected by stability; thus, in principle, a weak dependency on stability would be expected.

Lastly, we compare the frequency distribution of $z_0$ calculated with the analytical method during each season (Figure 4) from the IMPOWR campaign. Like for atmospheric stability, we find that surface roughness length is relatively insensitive to seasonal changes, with the mean and median values varying between $1.94 \times 10^{-3}$ m and $3.17 \times 10^{-3}$ m and between $4.16 \times 10^{-5}$ m and $1.04 \times 10^{-4}$ m, respectively (Table 4). The mean values are higher than the median values by an order of magnitude or more. The means from the Charnock method tend to be at least an order of magnitude lower than those from the other two methods. Lastly, the means from CW Historical tend to be larger than those from IMPOWR, especially with the analytical method (by at least an order of magnitude), possibly due to the higher reference height (20 vs. 12 m).

In conclusion, the median appears to be a more reliable and robust statistics than the mean for $z_0$. In terms of the two methods discussed in this section (i.e., analytical versus Charnock), we cannot yet recommend one over the other. We will address this issue in section 4.3 by comparing the wind speed profiles obtained with the various methods to recommend which to use for $z_0$. But first, we present the results obtained with the statistical method in the next section because, despite giving unrealistic values of $z_0$ at times, the statistical method offers many advantages that will be assessed in Section 4.3.

### 4.2. A Closer Look at the Statistical Method.
Looking back at the frequency distribution of $z_0$ for the two campaigns obtained from the statistical method (Figure 2(e)), the shape is clearly different from that of the analytical and Charnock methods, with two peaks at the highest and lowest bins. Note that the first and last bins include all $z_0$ values that are lower than $10^{-8}$ m and greater than 1 m, respectively. The mean $z_0$ values are extremely large (Table 4), greater than 1 m in all seasons and stabilities, which is a value usually associated with urban areas with tall multifloor buildings. High values of mean $z_0$, around 0.1 m, were obtained also by [20], who used the wind speeds from the CW Historical campaign at two heights (20 and 60 m) to calculate surface roughness directly using Equation (6) with $z_R = 20$ m. The median is more reasonable and closer to that of the other two methods, although generally higher (Table 4), varying between $3.16 \times 10^{-5}$ m in the DJF to $4.21 \times 10^{-2}$ m in JJA. The excessively high mean values and the unconventional shape of the frequency distribution of $z_0$ confirm that the statistical method is not providing realistic values of $z_0$, but rather $z_0$ values that best fit the vertical profiles of wind speed. Nonetheless, here we investigate why such unrealistic values are generated by the statistical method and whether or not it should still be used for offshore wind estimates.

TABLE 4: Statistical properties of surface roughness length during the CW Historical (2003–2007) and IMPOWR (2013-2014) campaigns at the Cape Wind tower obtained with the analytical, Charnock, and statistical methods. Note that the statistical method is not intended to give realistic values of surface roughness length, which are listed here in the last column just for completeness.

| | IMPOWR | | CW Historical | | |
| --- | --- | --- | --- | --- | --- |
| | Analytical | Charnock | Analytical | Charnock | Statistical |
| **ALL** | | | | | |
| Mean (m) | $2.36 \times 10^{-3}$ | $1.23 \times 10^{-4}$ | $1.13 \times 10^{-2}$ | $1.70 \times 10^{-4}$ | >1 |
| Median (m) | $5.98 \times 10^{-5}$ | $7.94 \times 10^{-5}$ | $6.19 \times 10^{-5}$ | $7.73 \times 10^{-5}$ | $6.09 \times 10^{-3}$ |
| Count | 18,764 | 18,819 | 99,309 | 100,393 | 207,676 |
| **DJF** | | | | | |
| Mean (m) | $3.00 \times 10^{-3}$ | $1.43 \times 10^{-4}$ | $1.17 \times 10^{-2}$ | $2.15 \times 10^{-4}$ | >1 |
| Median (m) | $9.12 \times 10^{-5}$ | $9.14 \times 10^{-5}$ | $7.25 \times 10^{-5}$ | $9.72 \times 10^{-5}$ | $3.16 \times 10^{-5}$ |
| Count | 3,257 | 3,263 | 23,066 | 23,309 | 49,213 |
| **MAM** | | | | | |
| Mean (m) | $1.94 \times 10^{-3}$ | $1.17 \times 10^{-4}$ | $1.47 \times 10^{-2}$ | $1.88 \times 10^{-4}$ | >1 |
| Median (m) | $4.30 \times 10^{-5}$ | $7.25 \times 10^{-5}$ | $3.82 \times 10^{-5}$ | $6.92 \times 10^{-5}$ | $3.9 \times 10^{-2}$ |
| Count | 9,559 | 9,594 | 26,479 | 26,931 | 49,876 |
| **JJA** | | | | | |
| Mean (m) | $3.17 \times 10^{-3}$ | $7.79 \times 10^{-5}$ | $0.91 \times 10^{-2}$ | $1.01 \times 10^{-4}$ | >1 |
| Median (m) | $4.16 \times 10^{-5}$ | $6.24 \times 10^{-5}$ | $5.22 \times 10^{-5}$ | $5.91 \times 10^{-5}$ | $4.21 \times 10^{-2}$ |
| Count | 2,266 | 2,276 | 15,650 | 15,871 | 52,860 |
| **SON** | | | | | |
| Mean (m) | $2.42 \times 10^{-3}$ | $1.51 \times 10^{-4}$ | $0.95 \times 10^{-2}$ | $1.56 \times 10^{-4}$ | >1 |
| Median (m) | $1.04 \times 10^{-4}$ | $1.01 \times 10^{-4}$ | $8.33 \times 10^{-5}$ | $8.26 \times 10^{-5}$ | $2.45 \times 10^{-4}$ |
| Count | 3,682 | 3,686 | 34,114 | 34,283 | 55,727 |
| **STABLE** | | | | | |
| Mean (m) | $3.74 \times 10^{-3}$ | $8.61 \times 10^{-5}$ | $1.69 \times 10^{-2}$ | $1.36 \times 10^{-4}$ | >1 |
| Median (m) | $2.96 \times 10^{-5}$ | $5.25 \times 10^{-5}$ | $1.67 \times 10^{-5}$ | $5.25 \times 10^{-5}$ | $1.03 \times 10^{-1}$ |
| Count | 7,339 | 7,006 | 24,687 | 25,536 | 25,093 |
| **NEUTRAL** | | | | | |
| Mean (m) | $1.11 \times 10^{-3}$ | $1.93 \times 10^{-4}$ | $1.29 \times 10^{-2}$ | $2.29 \times 10^{-4}$ | >1 |
| Median (m) | $1.33 \times 10^{-4}$ | $1.33 \times 10^{-4}$ | $3.94 \times 10^{-5}$ | $6.05 \times 10^{-5}$ | $3.2 \times 10^{-3}$ |
| Count | 1,776 | 2,882 | 7,567 | 7,625 | 7,346 |
| **UNSTABLE** | | | | | |
| Mean | $1.55 \times 10^{-3}$ | $1.39 \times 10^{-4}$ | $9.11 \times 10^{-3}$ | $1.76 \times 10^{-4}$ | >1 |
| Median | $8.69 \times 10^{-5}$ | $9.87 \times 10^{-5}$ | $1.00 \times 10^{-4}$ | $9.00 \times 10^{-5}$ | $1.88 \times 10^{-6}$ |
| Count | 9,649 | 8,931 | 67,055 | 67,228 | 62,292 |

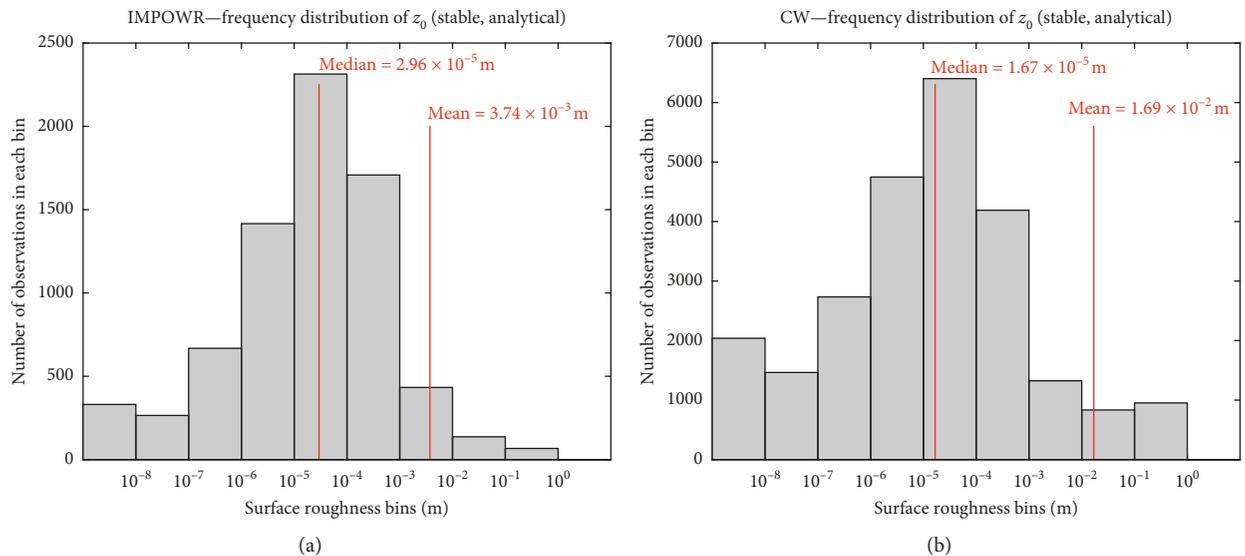

(a)  (b)

FIGURE 3: Continued.

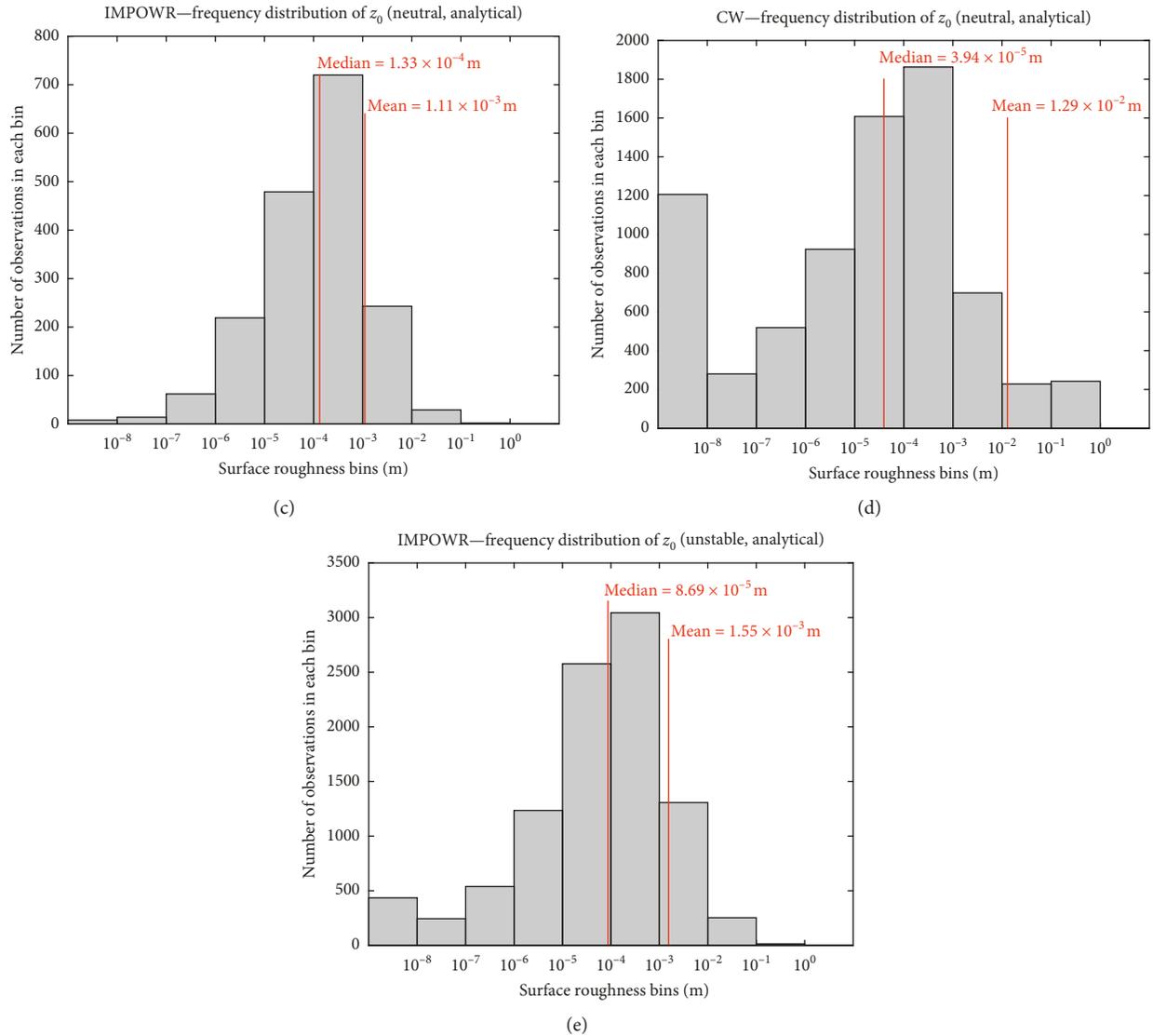

Figure 3: Frequency distribution of surface roughness length during the 2013–2014 IMPOWR campaign using two sonic anemometers located at 12 m AMSL (left) and during the 2003–2007 historical Cape Wind campaign (right) for (a, b) stable, (c, d) neutral, and (e, f) unstable cases. The mean and median $z_0$ are displayed on each figure. The $x$-axis is logarithmic and the $y$-axis has different maxima in each figure.

The reason why the frequency distribution with the statistical method is so different from that of the other two methods is that the statistical method predicts very high or very low values of $z_0$ when the vertical profile of wind speed is nonmonotonic. Whereas the typical wind speed profile is logarithmic and monotonic, Cape Wind is characterized by frequent nonmonotonic profiles [21], such that wind speed does not follow the log-law or does not increase with height. As suggested by [21], the nonmonotonic wind speed profiles are those that do not meet the monotonic condition:

$$U(20) < U(41) < U(60). \tag{11}$$

We apply the monotonic condition to the CW Historical data set and find that approximately 41% of the cases do not meet this condition and are therefore classified as non-monotonic profiles. When we apply the statistical method to just the monotonic cases (approximately 59%), the resulting mean and median $z_0$ values are closer to those obtained with the analytical and Charnock methods.

Because the nonmonotonic profiles occur too frequently to be ignored at Cape Wind, we want to retain such cases and try to understand why they give rise to unrealistic estimates of $z_0$. More importantly, we wonder if such unrealistic estimates of $z_0$ could actually be valuable still. We identify three types of nonmonotonic profiles. The first type (Case 1), an example of which is shown in Figure 5(a), has a zig-zag in the wind speed profile but enough shear that a log-like profile can be fit through the three points. The statistical method performs well in Case 1 and delivers a log-profile with realistic $z_0$ values; Case 1 occurs approximately 8% of the time. The second type (Case 2, Figure 5(b)), is basically shearless, i.e., the wind speed is very similar at all three levels. The only way to fit a log-like curve through a shearless profile is via a very low value of $z_0$. Shearless profiles occur ≈22% of the time at CW and the mean and median of $z_0$ are 2.08 × $10^{-8}$ m and 5.04 × $10^{-19}$ m, respectively, which can be considered unrealistically low and yet give profiles that fit very well the observed, nearly uniform wind speeds. The third type (Case 3, Figure 5(c)) is characterized by a decrease in wind speed with height (i.e., negative shear), as would occur during a sea-breeze event, and it is found on ≈11% of the time. Again, in order to fit a log-like curve through such a decreasing profile, a very high value of $z_0$ is required. Values of $z_0$ of the order of 100 m or more, which would be considered unrealistic, are actually a mathematical necessity to generate a profile with decreasing wind speeds with height.

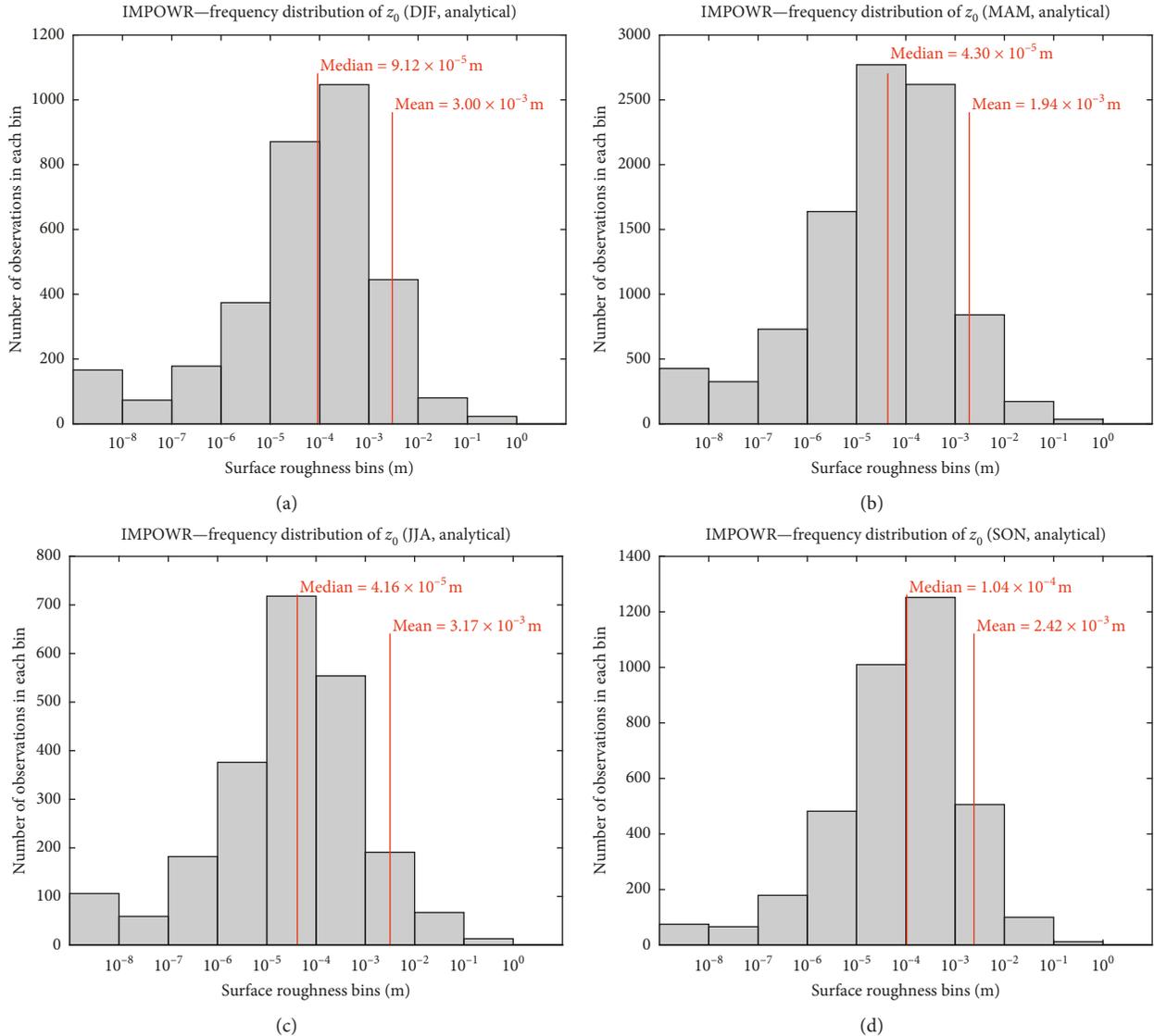

Figure 4: Frequency distribution of surface roughness length during the 2013–2014 IMPOWR campaign using two sonic anemometers located at 12 m AMSL for (a) December-January-February (DJF); (b) March-April-May (MAM); (c) June-July-August (JJA); and (d) September-October-November (SON). The mean and median $z_0$ are displayed on each figure. The x-axis is logarithmic and the y-axis has different maxima in each figure.

In summary, the statistical method does not always produce realistic estimates of $z_0$, but it always generates a profile that fits nicely within the three data points, even during cases characterized by nonmonotonic profiles, which are rather common at Cape Wind. As such, all the $z_0$ estimates generated by the statistical method should be retained and no minimum or maximum thresholds for the $z_0$ values should be imposed. As will be shown in the next section, the statistical method, despite its unrealistically high and low values of $z_0$ at times, generates the most accurate estimates of wind speed near hub height.

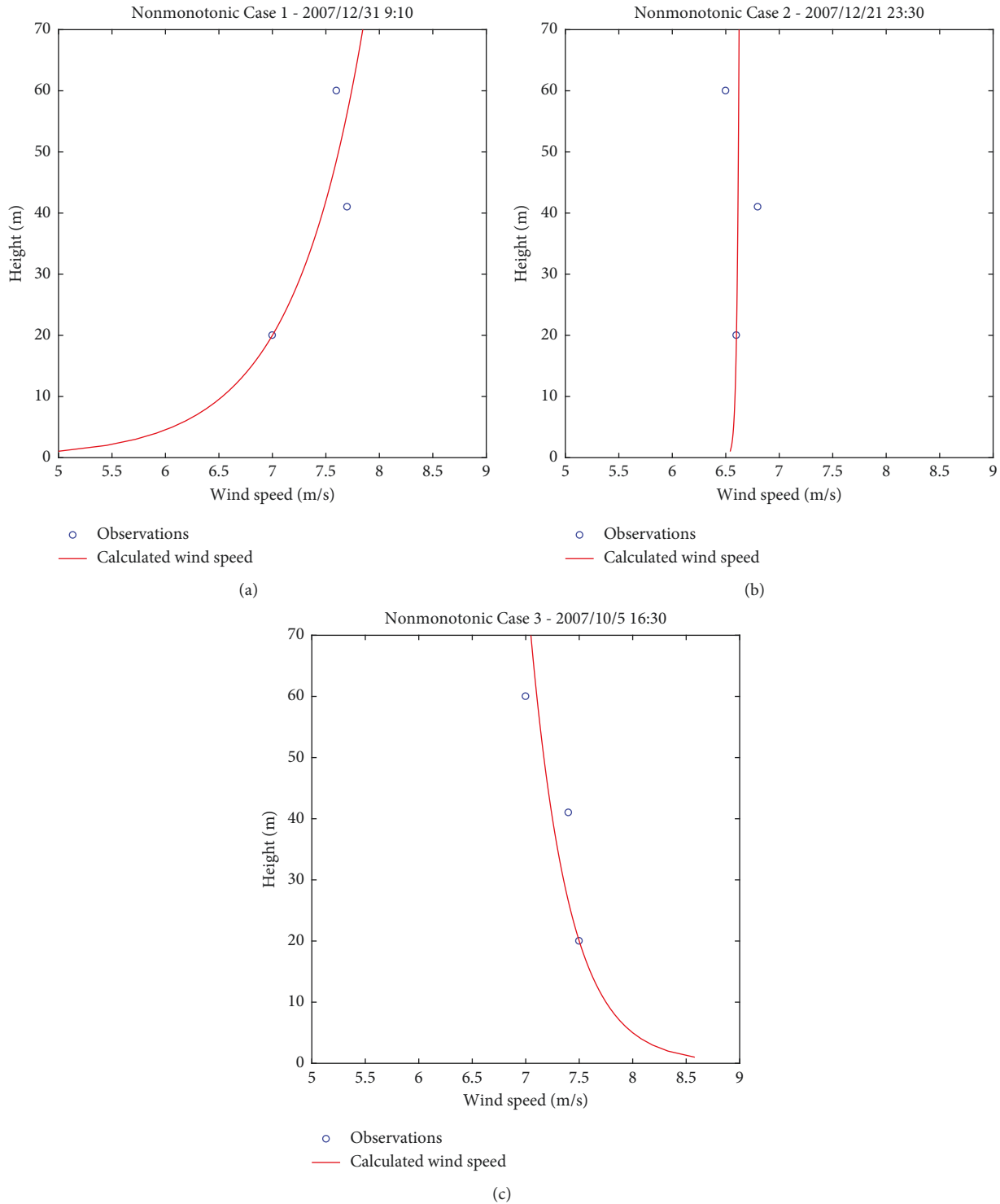

FIGURE 5: Examples of three types of observed, nonmonotonic wind speed profiles from the CW Historical data set and resulting log-fits using the statistical method: (a) Case 1 (zig-zag with positive wind shear and realistic $z_0$); (b) Case 2 (shearless with very low $z_0$); and (c) Case 3 (negative wind shear and very large $z_0$). The intercept of the log-fit (red line) and the $y$-axis represents the $z_0$ value in each figure.

### 4.3. Wind Speed Predictions near Hub Height.

Predicting the wind speed near hub height accurately is the ultimate application of surface roughness and the justification for developing fitting curves like the log-law. Here, we use the values of $z_0$ obtained via the three methods together with the wind speed observed at the reference height, 20 m for CW Historical and 12 m for IMPOWR, to estimate the wind speed at 60 m, the highest level of the CW tower. However, since the two campaigns were not simultaneous (CW Historical in 2003–2007 and IMPOWR in 2013–2014), a direct value-by-value comparison is impossible and therefore only average profiles will be compared. While the Charnock and analytical methods were applicable in both campaigns, the statistical method was only applicable to the CW Historical because IMPOWR did not have multilevel measurements; thus, a total of five curves are compared in Figure 6. The average observed wind speeds at the four available levels (12, 20, 41, and 60 m) are shown as circles and the fitting profiles with lines, solid for CW Historical and dashed for IMPOWR.

Using the 60 m level as the target, we compare the performance of the three methods next. Equation (4) is applied each time an observation at the reference height is available to generate an estimate of $U(z)$, using the value of $z_0$ that had been estimated for that same time via either the Charnock or the analytical methods. Similarly, Equation (6) is applied to the CW Historical data only, using the value of $z_0$ that had been estimated for that same time via the statistical method. At the end, the profiles shown in Figure 6 are obtained by averaging all the wind speed estimates for the entire period at each level $z$, after applying the monthly correction from Equation (7).

The first finding is that all methods give satisfactory wind speed profiles at Cape Wind, with average biases lower than 1 m/s at 60 m.

Among the three methods, we first compare the analytical and Charnock methods because both use only one-level measurements. At CW Historical, they both perform very well and their resulting vertical profiles are accurate and basically indistinguishable from each other. At IMPOWR, however, the analytical method outperforms Charnock's, as the resulting profiles on average are closer to the observations when $z_0$ is estimated with the analytical method (red dashed line in Figure 6). The physics-based properties of the analytical method and its slightly better performance over Charnock's were the reasons why the discussion of the seasonality of $z_0$ in the previous section (Figure 4) focused on the analytical method. The analytical method is thus recommended for projects with advanced sonic anemometry available at only one level.

The wind profiles predicted using $z_0$ from the Charnock method have the highest negative biases in the IMPOWR campaign, especially at the 60 m level. As such, Charnock is not the recommended method for estimating $z_0$ in offshore environments in which the purpose is to estimate hub-height wind speeds.

The statistical method, among the three methods, gives the most accurate estimate of the 60 m wind speeds (Figure 6). However, since the statistical method applied here actually utilizes the observed wind speeds at 60 m to estimate $z_0$, it is not surprising that it performs so well at 60 m. Before we can recommend the use of the statistical method for offshore wind energy applications, we need to perform an additional validation step. Thus, buoy 44020, located approximately 5 km to the northeast of the Cape Wind platform, is used to provide simultaneous wind speed observations at 5 m AMSL during the same period 2003–2007 as CW Historical. Using the buoy data, the statistical method is applied one more time to the CW Historical data set to estimate $z_0$ without using the 60-m wind speeds but using those at 5, 20, and 41 m in Equation (10). Next, the new $z_0$ values are used in Equation (6) to estimate $U(z)$ and the wind speed profiles are calculated over the whole period and averaged using Equation (7). In principle, each of the three levels can be used as the reference. Figure 7 shows that all three give an accurate estimate of the 60 m wind speed, but the 20 m reference height gives the best performance with a negligible positive bias of 0.03 m/s.

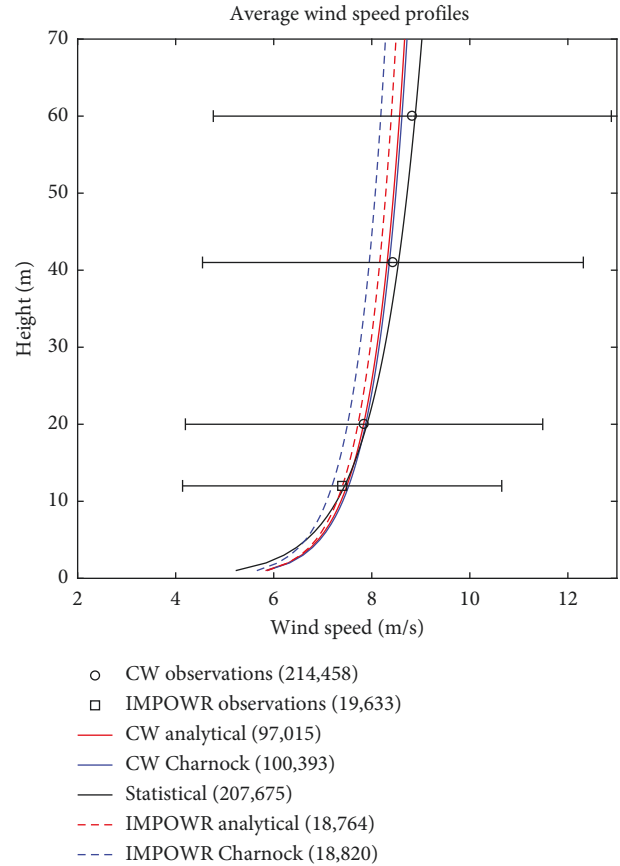

FIGURE 6: Average wind speed profiles calculated with $z_0$ values obtained with the three methods (Charnock, analytical, and statistical, in blue, red, and green) for both the CW Historical and IMPOWR campaigns (solid and dashed lines). The observations are displayed with solid circles.

We conclude therefore that the statistical method, despite nonphysical estimates of $z_0$ at times, gives such accurate predictions near hub height that it is recommended when multiple level measurements of wind speed are available with no stability information, as is the case in the CW Historical campaign with 20 m as the reference height or with vertically pointing floating lidars.

### 4.4. Can We Use a Single, Constant Value of Surface Roughness? The two recommended methods to calculate $z_0$, i.e., analytical and statistical, give excellent estimates of wind

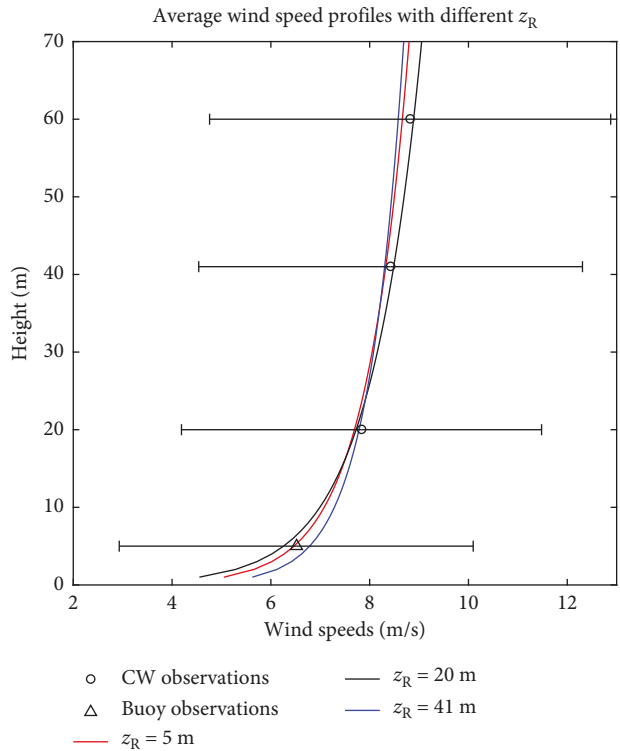

FIGURE 7: Average wind speed profiles obtained with $z_0$ estimates via the statistical method using 5 m wind speeds from buoy 44022 and 20 and 41 m from the CW Historical campaign to predict the wind speed at 60 m. Each of the three levels was tried as reference (5 m in red, 20 m in blue, and 41 m in green), and the best performance was with the 20 m level as reference.

speed near hub height in the offshore environment at Cape Wind (Figure 6). However, either one sophisticated 3D sonic anemometer or a multilevel tower equipped with cup anemometers are needed to calculate $z_0$ (Table 3). Here, we assess the validity of using a single, constant value of $z_0$ to estimate the near hub-height wind resource with limited data availability, namely, just wind speed at one level close to the surface. This is the case when only buoy data are available or when a numerical weather prediction model is used, but with coarse vertical resolution.

The next question is which value to pick. Given that, in Section 4.1, we concluded that the median value of $z_0$ is a more consistent and reliable statistics to characterize roughness than the mean, it makes sense to use a median. The first value that we test is $5.98 \times 10^{-5}$ m, which is the median from the analytical method in the IMPOWR campaign (Table 4). The statistical method resulted in a median $z_0$ value that was higher; thus, the second value that we test is $6.09 \times 10^{-3}$ m from the CW Historical campaign (Table 4). Since the median $z_0$ values obtained from the Charnock method in both campaigns ($7.94 \times 10^{-5}$ m and $7.73 \times 10^{-5}$ m) are very close to the median value from the analytical method used in this section ($5.98 \times 10^{-5}$ m), for simplicity they are not used in Figure 6.

The equation used to calculate the wind speed profiles with the two $z_0$ median values is the simplified log-law (Equation (6)) because it does not require stability information, which is not used in the statistical method and because we wanted to use the same equation with both methods. The simplified log-law equation only requires a reference height and, if a constant $z_0$ value is employed, the average observed wind speed at the reference height. Given that, we can use 5, 12, or 20 m as the reference height (buoy, IMPOWR, and CW Historical heights, respectively) and we have two values of $z_0$ to test; there are a total of six wind profiles in Figure 8. Whether we look at the 60 m level as the target or we consider the overall shape of the fitting profiles and how close they get to the observations, the three curves obtained with $z_0 = 6.09 \times 10^{-3}$ m (value obtained with the statistical method from CW Historical) give the best results regardless of the reference height (solid lines). The curves obtained with $z_0 = 5.98 \times 10^{-5}$ m (value obtained with the analytical method from IMPOWR) give reasonable results, except if the 5 m height is used a reference (red dashed line in Figure 8). The lower performance of the median $z_0 = 5.98 \times 10^{-5}$ m, combined with the fact that a value of $z_0$ of the order of $10^{-5}$ m is traditionally indicative of a very smooth surface, ice, or calm sea [33–35], lead us to recommend $z_0 = 6.09 \times 10^{-3}$ m as representative of this offshore region.

## 5. Conclusions

The main goal of this study was to provide accurate estimates and climatological properties of surface roughness length $z_0$ in the Nantucket Sound region, as well as more general recommendations for which of three methods (the Charnock and analytical, which are physical, and the statistical, which is mathematical), should be used to predict $z_0$ and ultimately wind speeds near hub height for future offshore wind farm development. Two field campaigns that were conducted at the Cape Wind (CW) platform, CW Historical in 2003–2007 and IMPOWR in 2013–2014, provided the observational data sets for the analysis.

The Nantucket Sound area is possibly a peculiar location due to the high frequency of nonmonotonic wind speed profiles, such as shearless wind profiles (12% of the cases) and profiles with negative shear (27%). These cases are not well represented by the analytical or Charnock methods, even when stability corrections are added, because the basic fit is always logarithmic and therefore monotonically increasing. By contrast, the statistical method represents them well, although it gives rise to unrealistic values of $z_0$ (very small and very large, respectively, for the shearless and negative-shear profiles).

Surface roughness length was found to be rather insensitive to seasonal changes and atmospheric stability, with mean and median $z_0$ values varying in the ranges $10^{-4} - 10^{-2}$ m and $10^{-5} - 10^{-4}$ m, respectively. Between mean and median, the median $z_0$ value is more consistent among the three different methods and the two campaigns. Among three methods analyzed in this study, the analytical and statistical methods resulted in the best wind speed estimates near hub height (i.e., at 60 m). However, the median $z_0$ values from the two methods are different by two orders of magnitude: $\approx 6 \times 10^{-5}$ m from the analytical method at IMPOWR, which is representative of very smooth surfaces, such as ice and calm seas, and $6.1 \times 10^{-3}$ m from the statistical method based on CW Historical data, more typical of rough seas. Which method should be picked? Consistent with [36], who concluded that no specific method can be proposed as the standard approach to calculate $z_0$, we too find that the best method depends on the type of observations that will be collected. Also, considering only $z_0$ values that are physical or consistent with traditional understanding from textbooks may be a limitation.

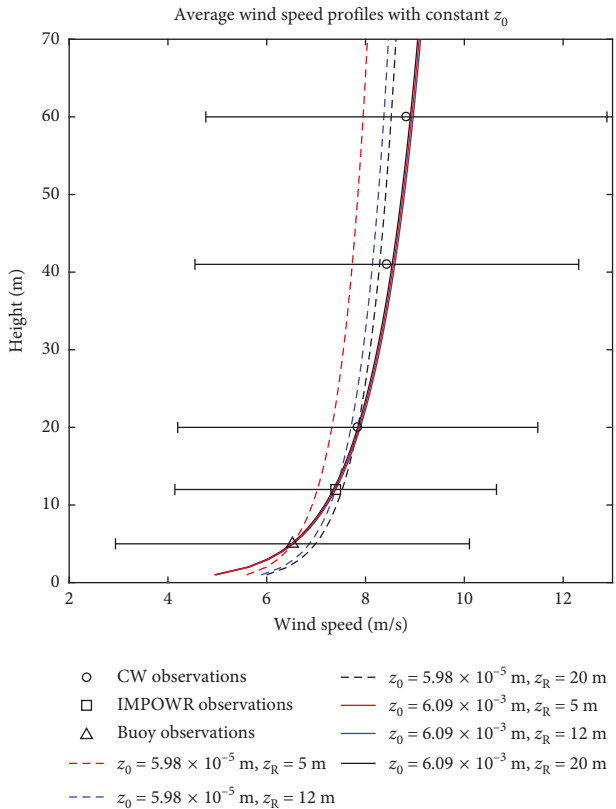

Figure 8: Average wind speed profiles based on two median values for $z_0$ obtained via the analytical method in the IMPOWR campaign (12 m reference height, $5.98 \times 10^{-5}$ m dashed lines) and via the statistical method in the CW Historical campaign (20-m reference height, $6.09 \times 10^{-3}$ m solid lines). The three reference heights are 5 m (typical buoy height, red), 12 m (blue), and 20 m (black).

If wind speed data will be collected at multiple levels above the water, as with meteorological towers on offshore platforms or with vertically pointing floating lidars, the statistical method is recommended to estimate $z_0$, even though it may provide unrealistic or nonphysical values of $z_0$ at times, because it will always generate a profile that is within the data bounds and accounts properly for nonmonotonic wind speed profiles. In a sense, the values of $z_0$ from the statistical method do not always match the traditional meaning of surface roughness because they include indirectly the effect of stability and nonmonotonic wind speed profiles.

If advanced 3D anemometry will be used at one level, then stability and $u_*$ information will be available and therefore this information can be effectively used in the analytical method, which will provide estimates of $z_0$ that are within the bounds of conventional understanding of surface roughness. The Charnock method too could be used in this case because $u_*$ is provided by the 3D sonic anemometer, but here we found that it gives estimates of near hub-height wind speeds that have a negative bias of about 1 m/s. Therefore, the Charnock method is not recommended to estimate $z_0$ for offshore wind applications.

If only wind speed measurements at one level are available, as with buoys or model outputs at coarse vertical resolution, we recommend the single, constant value of $6.1 \times 10^{-3}$ m, which is the median $z_0$ value from the statistical method at CW Historical, representative of rough seas.

## Data Availability

## Conflicts of Interest

The authors declare that they have no conflicts of interest.

## Acknowledgments

The authors thank Molly Kerrigan for her help at the initial stages of the research. Funding sources include the School of Marine Science and Policy Program Fellowship at the University of Delaware and the Delaware Natural Resources and Environmental Control (DNREC, award no. 18A00378).

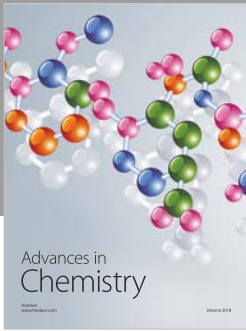
Advances in
Chemistry

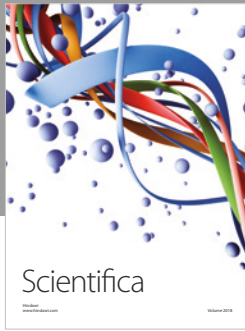
Scientifica

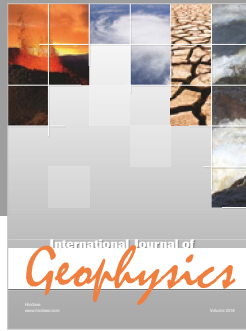
International Journal of
Geophysics

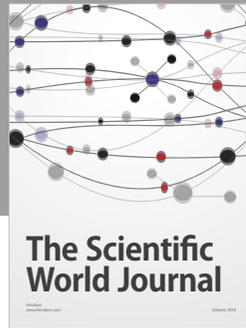
The Scientific World Journal

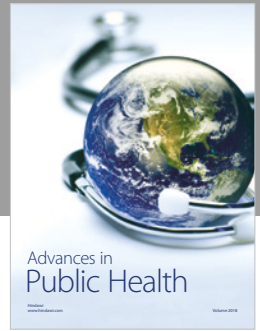
Advances in
Public Health

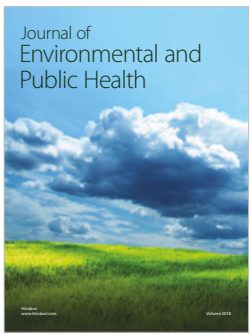
Journal of
Environmental and
Public Health

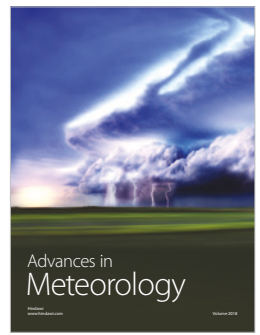
Advances in
Meteorology

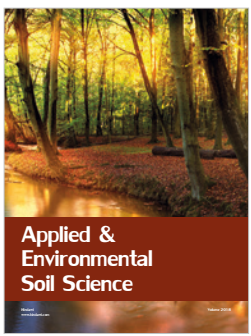
Applied &
Environmental
Soil Science

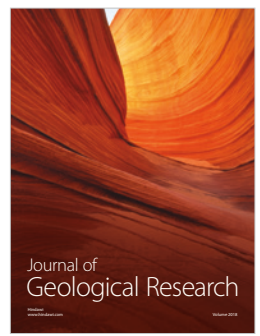
Journal of
Geological Research

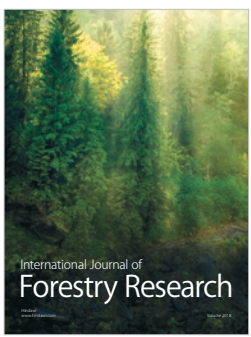
International Journal of
Forestry Research

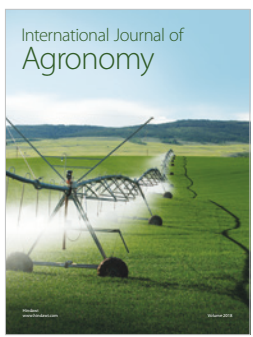
International Journal of
Agronomy

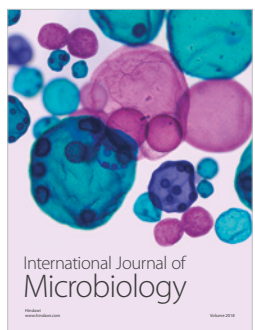
International Journal of
Microbiology

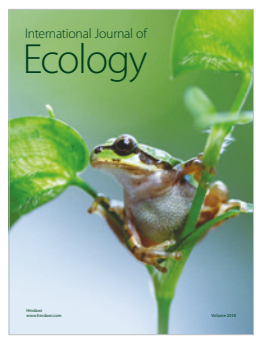
International Journal of
Ecology

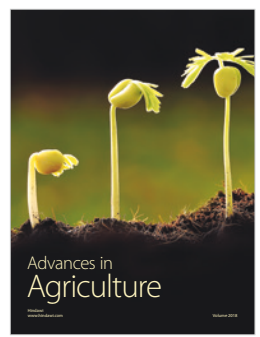
Advances in
Agriculture

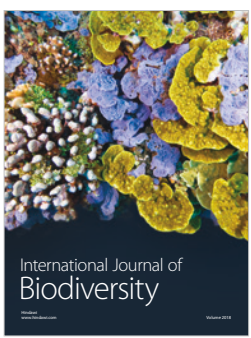
International Journal of
Biodiversity

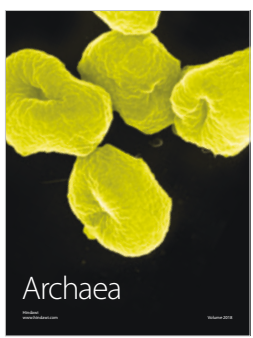
Archaea

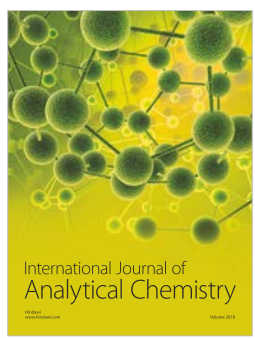
International Journal of
Analytical Chemistry

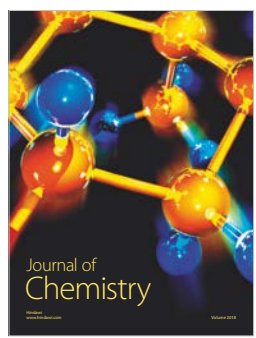
Journal of
Chemistry

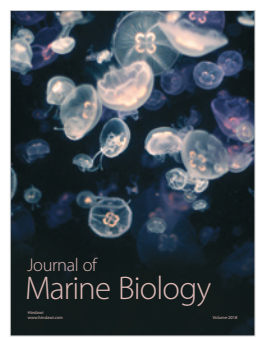
Journal of
Marine Biology

Hindawi

Submit your manuscripts at
www.hindawi.com